\documentclass[%
 reprint,
 superscriptaddress,
 amsmath,amssymb,
 aps,
 prl,
floatfix,
]{revtex4-1}

\usepackage{graphicx}
\usepackage{dcolumn}
\usepackage{bm}

\begin{document}

\preprint{APS/123-QED}

\author{Ingrid de Almeida Ribeiro}
\affiliation{Instituto de F\'{i}sica ``Gleb Wataghin'', Universidade Estadual de Campinas, UNICAMP, 13083-859, Campinas, S\~{a}o Paulo, Brazil}
\email{iribeiro@ifi.unicamp.br}
\author{Maurice de Koning}
\affiliation{Instituto de F\'{i}sica ``Gleb Wataghin'', Universidade Estadual de Campinas, UNICAMP, 13083-859, Campinas, S\~{a}o Paulo, Brazil}
\affiliation{Center for Computing in Engineering \& Sciences, Universidade Estadual de Campinas, UNICAMP, 13083-861, Campinas, S\~{a}o Paulo, Brazil}
\email{dekoning@ifi.unicamp.br}

\date{\today}

\title
{Non-Newtonian flow effects in supercooled water}

\begin{abstract}

The viscosity of supercooled water has been a subject of intense study, in particular with respect to its temperature dependence. Much less is known, however, about the influence of dynamical effects on the viscosity in its supercooled state. Here we address this issue for the first time, using molecular dynamics simulations to investigate the shear-rate dependence of the viscosity of supercooled water as described by the TIP4P/Ice model. We show the existence of a distinct cross-over from Newtonian to non-Newtonian behavior characterized by a power-law shear-thinning regime. The viscosity reduction is due to the decrease in the connectivity of the hydrogen-bond network. Moreover, the shear thinning intensifies as the degree of supercooling increases, whereas the cross-over flow rate is approximately inversely proportional to the Newtonian viscosity. These results stimulate further investigation into possible fundamental relations between these nonequilibrium effects and the quasi-static Newtonian viscosity behavior of supercooled water.
\end{abstract}

\maketitle


Supercooled liquid water has been the subject of intense investigation for decades~\cite{Angell1983,Gallo2016} and continues to attract significant attention~\cite{Cerdeirina2019,Hestand2018,Naserifar2019}. Besides the hotly debated issue concerning the possible existence of a second critical point in the supercooled regime~\cite{Limmer2011,Limmer2013,Palmer2018}, there has been a long-standing interest in the behavior of water's viscosity below the melting temperature. Particular topics of interest include the existence of a fragile-to-strong transition~\cite{Ito1999,Shi2018}, the relation between viscosity and molecular diffusion~\cite{Dehaoui2015} and the effect of pressure~\cite{Singh2017}. 

The viscosity $\eta$ of a viscous fluid is defined as the proportionality constant between the shear stress $\sigma$ and the corresponding strain rate $\dot{\gamma}$ according to $\sigma = \eta \, \dot{\gamma}$~\cite{Lakes2009,Krishnan2010}. If, for given temperature and pressure, the relation between $\sigma$ and $\dot{\gamma}$ is linear, i.e., $\eta$ is constant, the flow behavior of the fluid is said to be Newtonian~\cite{Krishnan2010}. Conversely, fluids for which this linearity is violated are referred to as non-Newtonian, with colloidal suspensions, many polymer melts and granular fluids as typical examples~\cite{Larson1999,Krishnan2010}. 

Many fluids display Newtonian flow behavior for sufficiently small rates $\dot{\gamma}$. Liquid water in thermodynamic equilibrium is an example, with a viscosity that is known to be constant across several orders of magnitude of $\dot{\gamma}$~\cite{Khatibi2018}. Much less is known, however, about the dynamical effects on the viscosity of water in its supercooled state. Although its magnitude is known to rise sharply as the temperature is lowered~\cite{Hallett1963,Dehaoui2015}, this increase has so far only been probed for the low-rate, Newtonian limit and the question as to whether it displays a shear-rate dependence remains open. 

In this Letter we consider this issue for the first time, investigating the influence of the flow-rate on the shear viscosity of supercooled water using atomistic-level simulations. In particular, we employ non-equilibrium molecular dynamics (NEMD) simulations in which we impose shear deformations at a constant rate $\dot{\gamma}$ and measure the associated shear stress $\sigma$. To describe the interactions between the water molecules we employ the TIP4P/Ice water model~\cite{Abascal2005}, which is among the best molecular models for water~\cite{Haji-Akbari2017} and has a melting point $T_m=271\, K$ that is close to the experimental value. All simulations have been carried out using the \verb|LAMMPS| package~\cite{Plimpton1995}. The long-range intermolecular electrostatic interactions for the TIP4P/Ice model are calculated using the particle-particle particle-mesh (PPPM) scheme~\cite{Hockney2017} and the intramolecular bond lengths and angles are held fixed using the \verb|SHAKE| algorithm~\cite{Ryckaert1977}. 

All the flow simulations are carried out using a computational cell containing 10800 water molecules. The cells are first allowed to equilibrate at zero external pressure and constant temperature, allowing fully flexible cells. This is achieved using a Parrinello-Rahman-type barostat~\cite{Shinoda2004} and a Langevin thermostat~\cite{Schneider1978} with damping constants of 2 and 0.2 ps, respectively. The corresponding equations of motion are integrated using velocity-Verlet algorithm with a time step of $\Delta t = 1$ fs. Subsequently, the nonequilibrium flow simulations are carried out at constant volume and isothermally, with temperature control implemented using a Langevin thermostat with a damping constant of 0.2 ps. The pure shear deformations are imposed using \verb|LAMMPS|'s \verb|fix deform| command with the \verb|remap x| option,  allowing the molecules to adjust to the cell deformation without requiring an explicit velocity profile. This approach has shown to give good agreement with the alternative SLLOD approach~\cite{Hagita2017}. Due to the appreciable cell distortions during the NEMD simulations, the reciprocal space part of the PPPM scheme is reset several times during a run, approximately after every $\sim 1$\% of deformation. 

\begin{figure}[ht!]
\includegraphics[width=8.4cm]{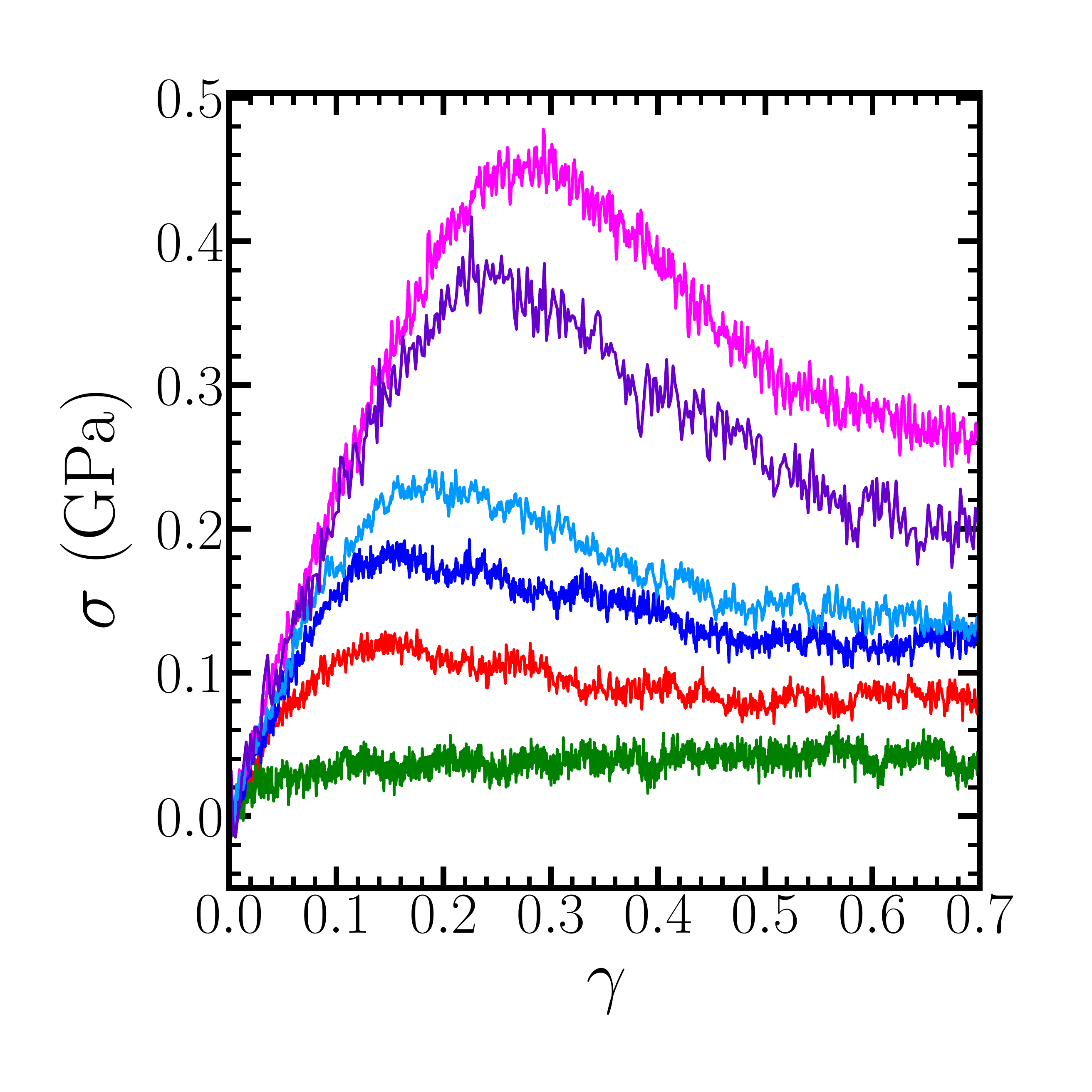}\hfill
\caption{\label{fig1} Shear stress as a function of the accumulated strain at $T=226\, K$ for $\dot{\gamma}=2\times10^7$ s$^{-1}$ (green), $2\times10^8$ s$^{-1}$ (red), $1\times 10^9$ s$^{-1}$ (dark blue), $2.5\times 10^9$ s$^{-1}$ (light blue), $2.5\times 10^{10}$ s$^{-1}$ (purple), and $5\times 10^{10}$ s$^{-1}$ (magenta).}
\end{figure}

Fig.~\ref{fig1} displays the evolution of the shear stress as a function of the accumulated strain, $\gamma = \dot{\gamma}\,t$, along six flow simulations at the deeply supercooled condition at $T=226\, K$. 

The stress-strain curves display non-monotonic behavior that is typical of viscoelastic fluids, as has been observed in a variety of systems, both experimentally as well as in simulations~\cite{Osaki2000,Islam2001,Letwimolnun2007,Varnik2004,Zausch2008,Zausch2009,Fuereder2017}. At the early stages of the flow process the stress increases linearly with strain, typifying a solid-like elastic response characterized by a modulus that is independent of the deformation rate. Subsequently, the contribution of viscous relaxation processes becomes significant, first reducing the elastic increase of the shear stress to reach a maximum, $\sigma_{\rm max}$, followed by a final decay to a steady-state plateau value, $\sigma_{\infty}$. Both $\sigma_{\rm max}$ and $\sigma_{\infty}$ decrease as the flow rate is reduced, as the stress relaxation processes are active during longer periods of time for a given state of deformation. Indeed, for $\dot{\gamma}=2\times10^7$ s$^{-1}$ the stress maximum has disappeared altogether and the stress-strain curve rises monotonically to its steady-state value.

\begin{figure*}[ht!]
\includegraphics[width=17cm]{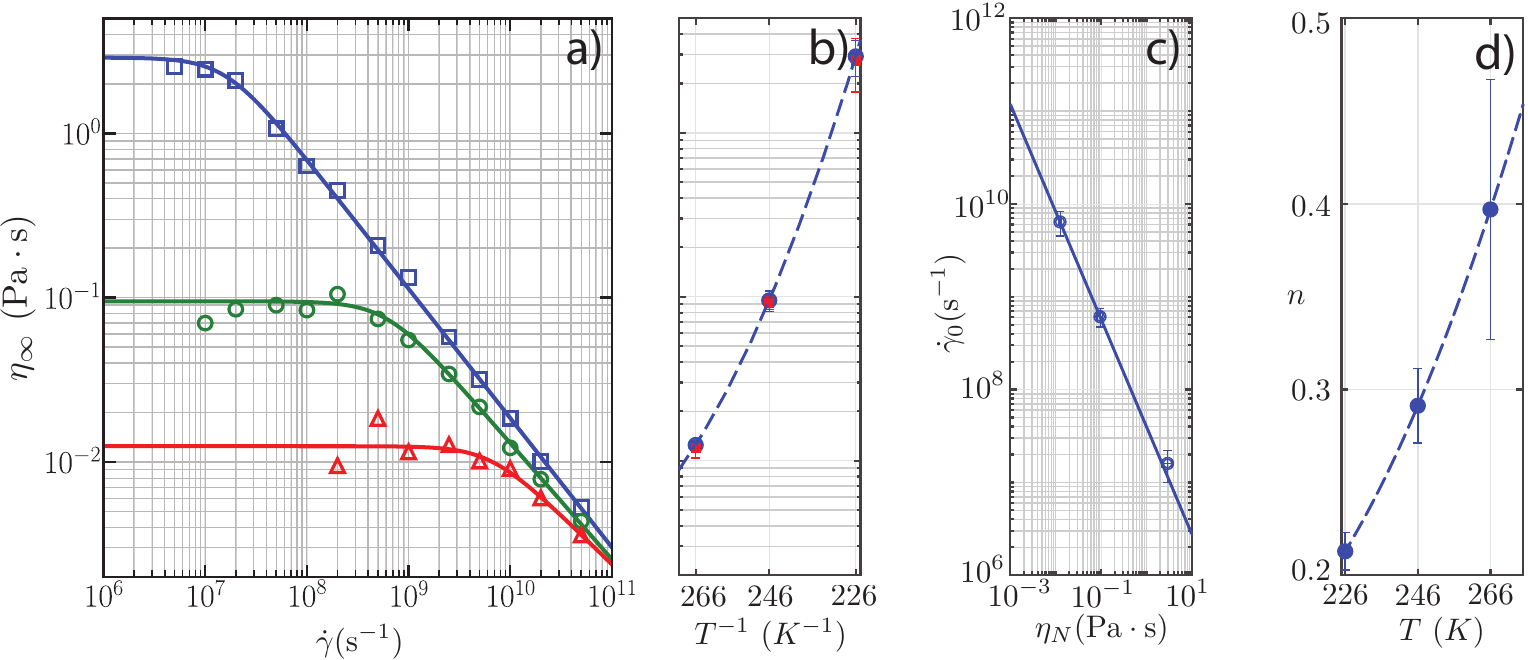}\hfill
\caption{\label{fig2} a) NEMD shear viscosity as a function of the flow rate for supercooled TIP4P/Ice water at $T=226$ (squares), 246 (circles) and $266\, K$ (triangles). Error bars are smaller than symbol size and are not shown. Solid lines correspond to fits of the viscosity data to the Carreau model, Eq.~(\ref{Carreau}). b) Comparison of Carreau estimate for $\eta_N$ (circles) to Green-Kubo results (squares), as a function of the inverse temperature $1/T$. c) Characteristic cross-over rate $\dot{\gamma}_0$ as a function of $\eta_N$. Full line represents power-law fit with exponent $-1.16 \pm 0.01$. d) Shear thinning exponent $n$ as a function of temperature $T$. Dashed lines in b) and d) represent guides to the eye. Error bars in b), c) and d) correspond to 95\% confidence intervals.}
\end{figure*}

The plateau value $\sigma_{\infty}$ is the shear stress that is required to maintain steady-state flow at a prescribed rate $\dot{\gamma}$ and the corresponding steady-state shear viscosity is then given by
\begin{equation}
\eta_{\infty}(\dot{\gamma})\equiv\sigma_{\infty}(\dot{\gamma})/\dot{\gamma}.
\end{equation}
Fig.~\ref{fig2}a) displays this viscosity as a function of flow rate for supercooled TIP4P/Ice water at 226 $K$, 246 $K$ and 266 $K$, respectively. For all three temperatures the flow response can be classified into two regimes. For low rates the viscosity is independent of $\dot{\gamma}$, meaning that flow is Newtonian under these conditions.  Subsequently, there is a cross-over into a non-Newtonian regime in which the viscosity decreases with growing flow rates, also known as shear thinning. Furthermore, this cross-over depends strongly on the temperature: while at $226 \, K$ non-Newtonian behavior sets in for $\dot{\gamma} \gtrsim 10^7$ s$^{-1}$, the Newtonian flow regime persists up to flow rates of $\dot{\gamma} \sim 10^{10}$ s$^{-1}$ at $266\, K$. 

To quantify the cross-over between Newtonian and non-Newtonian flow we analyze the simulation data in terms of the Carreau model~\cite{Carreau1972,Spikes2014,Valencia2019,Jadhao2017}, which provides a phenomenological description of shear thinning that has shown to be accurate for fluids with relatively low Newtonian viscosities, $\eta_N \lesssim 1$ Pa$\cdot$s~\cite{Jadhao2017}, which is the case for the present TIP4P/Ice simulations. The Carreau model treats shear flow as a stress-assisted thermally activated process involving a broad distribution of energy barriers and gives a shear viscosity that depends on the flow rate according to~\cite{Carreau1972,Spikes2014,Jadhao2017}
\begin{equation}
\label{Carreau}
\frac{\eta_{\infty}}{\eta_N}=\left[1+\left(\frac{\dot{\gamma}}{\dot{\gamma}_0}\right)^2\right]^{\frac{n-1}{2}},
\end{equation}
where $\eta_{N}$ is the Newtonian viscosity,  $\dot{\gamma_0}$ is a characteristic cross-over rate and $n$ is the shear-thinning exponent with a value between 0 and 1. In the limit of large flow rates this model gives rise to a power-law decay of the viscosity according to $\eta_{\infty} \sim \dot{\gamma}^{n-1}$.

The lines in Fig.~\ref{fig2}a) depict the least-squares regression results for the Carreau model of Eq.~(\ref{Carreau}) with respect to the NEMD viscosity data. The agreement between model and simulation is very good across the entire range of flow rates for all three temperatures, clearly showing a power-law dependence of the viscosity in the shear thinning regime. The accuracy of the Carreau model can be further verified by comparing its estimate for the Newtonian viscosity $\eta_N$ to results from independent equilibrium calculations. Specifically, since $\eta_N$ represents the shear viscosity in the limit of vanishing flow rate, it can be computed using the Green-Kubo (GK) formalism~\cite{Hansen2006,Allen2017,McQuarrie2000}, which expresses it in terms of stress-stress autocorrelation functions that can be computed using equilibrium MD simulations. The equilibrium runs used to compute the GK viscosities are based on a cubic cell containing 2000 water molecules that are first equilibrated at zero pressure and constant temperature using the same approach used for the 10800-molecule cells. Subsequently, five independent NVT equilibrium runs are carried out to sample the components of the stress tensor and determine the stress-stress autocorrelation functions $\langle\,  P_{\alpha \beta}(0) P_{\alpha \beta}(t) \,\rangle$, where $P_{\alpha \beta}$ is an off-diagonal component of the stress tensor. The Green-Kubo viscosities are then computed as
\begin{equation}
\label{eta}
\nonumber
\eta_N = \frac{V}{k_{B}T}\int_{0}^{\infty}\langle\,  P_{\alpha \beta}(0) P_{\alpha \beta}(t) \,\rangle dt,
\end{equation}
with $V$ the volume of the system, $T$ the temperature, and $k_{B}$ Boltzmann's constant. Aside from the three off-diagonal components $P_{xy}$, $P_{xz}$, and $P_{zy}$, there are two other independent components, $\frac{1}{2}(P_{xx}-P_{yy})$ and $\frac{1}{2}(P_{yy}-P_{zz})$, that can be used due to rotational invariance~\cite{Alfe1998}. Accordingly, $\eta_N$ is estimated using the average over these five components and over five independent equilibrium runs.

Fig.~\ref{fig2}b) presents a comparison between the NEMD Carreau results and equilibrium GK shear viscosities. The agreement is excellent for all three temperatures, providing further validation of the Carreau model as an adequate descriptor of the rate dependence of the shear viscosity in supercooled TIP4P/Ice water. A further observation based on the results in Fig.~\ref{fig2}b) is that supercooled TIP4P/Ice water behaves as a fragile liquid for the considered temperatures~\cite{Debenedetti2001}, given that the logarithm of $\eta_N$ as a function of the inverse temperature $1/T$ is supralinear, constituting super-Arrhenius behavior.

\begin{figure*}[ht!]
\includegraphics[width=17cm]{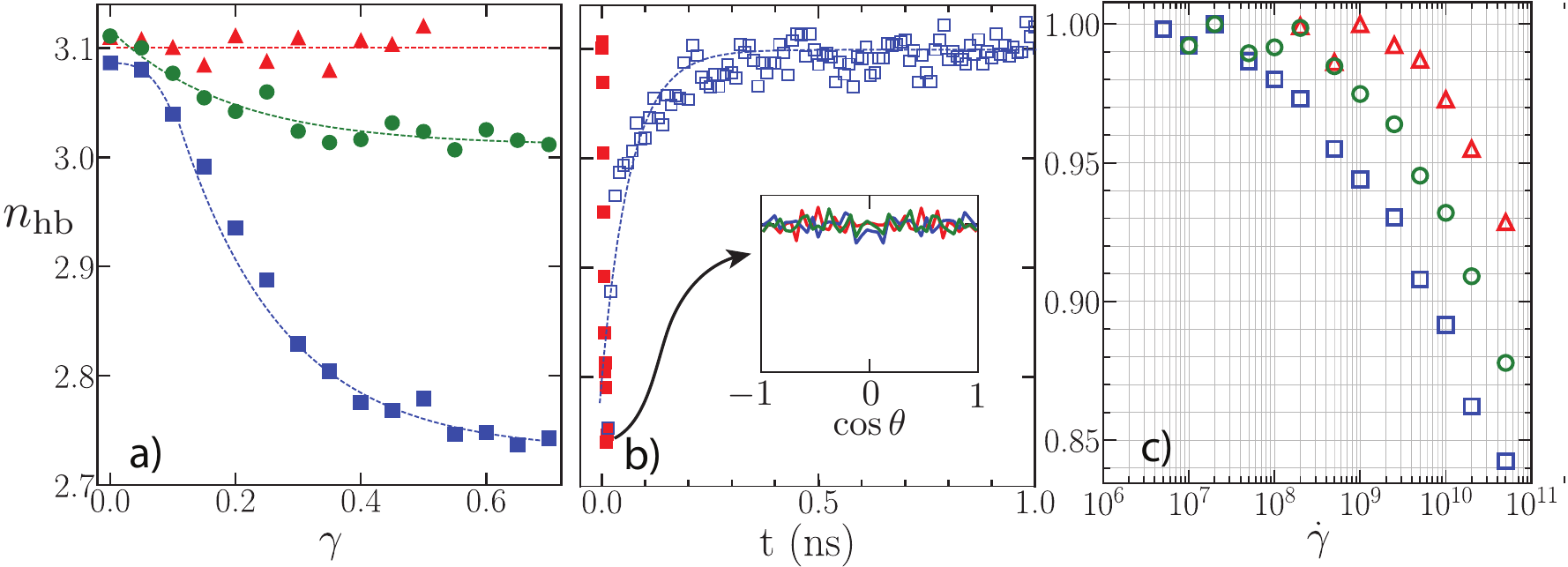}\hfill
\caption{\label{fig3} Average number of hydrogen bonds per molecule $n_{\rm hb}$ during flow simulations at $T=246~K$. (a) Results for $\dot{\gamma}=2\times 10^8$ (triangles), $2.5\times10^9$ (circles) and $5\times 10^{10}$ s$^{-1}$, respectively. Dashed lines serve as guides to the eye. (b) Temporal evolution of $n_{\rm hb}$ during a simulation in which the system is first subjected to a constant flow rate of $\dot{\gamma}=5\times 10^{10}$ s$^{-1}$ until reaching a total shear of $\gamma=0.7$ (filled squares), after which the deformation is instantaneously halted and the system is allowed to relax at a fixed cell geometry (open squares). Lines in inset display distribution of HB direction cosines with respect to $x$ (red), $y$ (blue) and $z$ (green) directions at $\gamma=0.7$.(c) Variation of $n_{\rm hb}$ normalized by its equilibrium value as a function of $\dot{\gamma}$.}
\end{figure*}

The two other parameters of the Carreau model quantify the nature of the Newtonian to non-Newtonian transition and their behavior is plotted in Figs.~\ref{fig2}c) and d). Fig.~\ref{fig2}c) plots the characteristic rate $\dot{\gamma}_0$ as a function of the Newtonian viscosity $\eta_N$. As noted before, the transition to the shear-thinning regime sets in for lower flow rates as the temperature reduces and the Newtonian viscosity grows. More interestingly, the functional dependence is well described by a power law with exponent $-1.16\pm 0.01$, implying a direct relationship between the nonequilibrium parameter $\dot{\gamma}_0$ and the equilibrium property $\eta_N$. Fig.~\ref{fig3}c) shows that the shear thinning exponent $n$ decreases substantially as the degree of supercooling is enhanced, implying that the shear-thinning effect becomes more pronounced as the temperature is reduced. We will further discuss this point below. 

There are a number of microscopic processes that can lead to the power-law viscosity behavior of the Carreau model seen in Fig.~\ref{fig2}a)~\cite{Jadhao2017}. A common mechanism concerns a change in some order parameter that describes correlations between neighboring molecules~\cite{Jadhao2017,Loose1989}. For instance, for shear thinning in fluids composed of chain molecules, a relevant order parameter is one that measures their alignment along the flow direction~\cite{Kroger1993,Petravic2005}. Here, we investigate the evolution of the hydrogen bonding during the flow simulations. 
To determine the hydrogen-bond statistics, we adopt the definition that a HB is present whenever the distance between a proton and an oxygen satisfies $1.1$ \AA $< d_{\mbox{OH}} < 2$ \AA. Fig.~\ref{fig3}a) displays the mean number of hydrogen bonds (HBs) per molecule, $n_{\rm hb}$, as a function of strain at $T=246~K$ for the flow rates $\dot{\gamma} = 2\times 10^8$,  $2.5\times10^9$ and $5\times 10^{10}$ s$^{-1}$. These particular three values correspond to the Newtonian, the cross-over and shear-thinning regimes for this temperature, respectively. In the Newtonian regime $n_{\rm hb}$ remains constant throughout the entire simulation and the connectivity of the HB network remains unaffected by the flow. As the rate increases to the Carreau cross-over value, however, the steady-state HB connectivity becomes discernibly lower, reducing even further for the highest flow rate. 

As mentioned above, molecular alignment during the shearing process may also possibly play a role in the shear thinning, as is the case in systems where elongated molecules are involved~\cite{Kroger1993,Petravic2005}. To verify this possibility for water we analyze the statistics of HB directions during the shearing process. As seen in the inset of Fig.~\ref{fig3}b), the HB direction cosines with respect to the $x$, $y$ and $z$ directions are uniformly distributed, indicating that the HB directionality is isotropic, displaying no preferred alignment direction. 

These results indicate that the shear thinning arises from the reduction of HB connectivity, which is consistent with theoretical arguments~\cite{Lubchenko2009}. The origin of this decrease and its dependence on the flow rate is associated with time-scale differences between the imposed flow and molecular rearrangements. In the Newtonian regime the latter is sufficiently short for the molecular rearrangements to accompany the imposed flow and maintain the average connectivity of the HB network. In the non-Newtonian shear-thinning regime this is no longer the case, with the molecular orientations systematically lagging behind the imposed flow, leading to the reduction of the HB connectivity in the steady state flow. This is illustrated in Fig.~\ref{fig3}b) which depicts the time evolution of $n_{\rm hb}$ along a simulation in which the system is first subjected to a constant flow rate of $\dot{\gamma}=5\times 10^{10}$ s$^{-1}$ at 246 $K$ until reaching its steady state, after which the deformation is halted and the system is allowed to relax at a fixed cell geometry. During the flow stage the mean number of HBs per molecule rapidly decreases to its steady-state value. Subsequently, after halting the deformation, $n_{\rm hb}$ relaxes to its equilibrium value by an approximately exponential process with a time constant $\tau_m \simeq 0.07$ ns.  While this time scale is $\sim 300$ times shorter than that associated with the lowest flow rate in Fig.~\ref{fig3}a), it is $\sim 3$ times larger compared to that of the highest. 

Finally, the increasing intensity of the shear-thinning effect with reducing temperature, as reflected by the decrease of the Carreau exponent $n$ in Fig.~\ref{fig2}d), also correlates with the evolution of the average number of $n_{\rm hb}$. This is shown in Fig.~\ref{fig3}c) which depicts the steady-state flow values of $n_{\rm hb}$, normalized by their equilibrium values $n^0_{\rm hb}$, as a function of the flow rate for $T=226$, 246 and 266~$K$, respectively. Due to the shear thinning effect, as seen in Fig.~\ref{fig2}a), $n_{hb}$ decreases as the flow rate grows. Moreover, this decrease is stronger in relative terms as the temperature is lowered: whereas for $\dot{\gamma}=5\times10^{10}$ s$^{-1}$ a reduction of $\sim 15$\% with respect to its equilibrium value is observed at 226 $K$, it is only $\sim 6.5$\% at 266 $K$.   

In conclusion, we have performed a series of NEMD simulations to investigate the shear-rate dependence of the viscosity of supercooled water as described by the TIP4P/Ice model for three different degrees of supercooling. In all cases we find a distinct Newtonian-to-shear-thinning crossover that is well-described by the Carreau model. The shear-thinning effect becomes stronger as the temperature is reduced, with a thinning exponent that decreases and with non-Newtonian behavior setting in for lower deformation rates. Interestingly, the results suggest a power-law relationship between the nonequilibrium cross-over rate parameter $\dot{\gamma}_0$ and the equilibrium Newtonian viscosity property $\eta_N$. On the molecular scale the shear thinning correlates with a significant reduction in the connectivity of the HB network, which is associated with time-scale differences between the deformation protocol and molecular rearrangements. Moreover, the connectivity reduction increases in relative terms as the temperature is lowered, giving rise to the stronger shear-thinning effect at lower temperatures. 

\acknowledgments
The authors acknowledge support from CNPq, Fapesp grant no. 2016/23891-6 and the Center for Computing in Engineering \& Sciences - Fapesp/Cepid no. 2013/08293-7.


\begin{thebibliography}{45}%
\makeatletter
\providecommand \@ifxundefined [1]{%
 \@ifx{#1\undefined}
}%
\providecommand \@ifnum [1]{%
 \ifnum #1\expandafter \@firstoftwo
 \else \expandafter \@secondoftwo
 \fi
}%
\providecommand \@ifx [1]{%
 \ifx #1\expandafter \@firstoftwo
 \else \expandafter \@secondoftwo
 \fi
}%
\providecommand \natexlab [1]{#1}%
\providecommand \enquote  [1]{``#1''}%
\providecommand \bibnamefont  [1]{#1}%
\providecommand \bibfnamefont [1]{#1}%
\providecommand \citenamefont [1]{#1}%
\providecommand \href@noop [0]{\@secondoftwo}%
\providecommand \href [0]{\begingroup \@sanitize@url \@href}%
\providecommand \@href[1]{\@@startlink{#1}\@@href}%
\providecommand \@@href[1]{\endgroup#1\@@endlink}%
\providecommand \@sanitize@url [0]{\catcode `\\12\catcode `\$12\catcode
  `\&12\catcode `\#12\catcode `\^12\catcode `\_12\catcode `\%12\relax}%
\providecommand \@@startlink[1]{}%
\providecommand \@@endlink[0]{}%
\providecommand \url  [0]{\begingroup\@sanitize@url \@url }%
\providecommand \@url [1]{\endgroup\@href {#1}{\urlprefix }}%
\providecommand \urlprefix  [0]{URL }%
\providecommand \Eprint [0]{\href }%
\providecommand \doibase [0]{http://dx.doi.org/}%
\providecommand \selectlanguage [0]{\@gobble}%
\providecommand \bibinfo  [0]{\@secondoftwo}%
\providecommand \bibfield  [0]{\@secondoftwo}%
\providecommand \translation [1]{[#1]}%
\providecommand \BibitemOpen [0]{}%
\providecommand \bibitemStop [0]{}%
\providecommand \bibitemNoStop [0]{.\EOS\space}%
\providecommand \EOS [0]{\spacefactor3000\relax}%
\providecommand \BibitemShut  [1]{\csname bibitem#1\endcsname}%
\let\auto@bib@innerbib\@empty
\bibitem [{\citenamefont {Angell}(1983)}]{Angell1983}%
  \BibitemOpen
  \bibfield  {author} {\bibinfo {author} {\bibfnamefont {C.~A.}\ \bibnamefont
  {Angell}},\ }\href {\doibase 10.1146/annurev.pc.34.100183.003113} {\bibfield
  {journal} {\bibinfo  {journal} {Annu. Rev. Phys. Chem.}\ }\textbf {\bibinfo
  {volume} {34}},\ \bibinfo {pages} {593} (\bibinfo {year} {1983})}\BibitemShut
  {NoStop}%
\bibitem [{\citenamefont {Gallo}\ \emph {et~al.}(2016)\citenamefont {Gallo},
  \citenamefont {Amann-Winkel}, \citenamefont {Angell}, \citenamefont
  {Anisimov}, \citenamefont {Caupin}, \citenamefont {Chakravarty},
  \citenamefont {Lascaris}, \citenamefont {Loerting}, \citenamefont
  {Panagiotopoulos}, \citenamefont {Russo}, \citenamefont {Sellberg},
  \citenamefont {Stanley}, \citenamefont {Tanaka}, \citenamefont {Vega},
  \citenamefont {Xu},\ and\ \citenamefont {Pettersson}}]{Gallo2016}%
  \BibitemOpen
  \bibfield  {author} {\bibinfo {author} {\bibfnamefont {P.}~\bibnamefont
  {Gallo}}, \bibinfo {author} {\bibfnamefont {K.}~\bibnamefont {Amann-Winkel}},
  \bibinfo {author} {\bibfnamefont {C.~A.}\ \bibnamefont {Angell}}, \bibinfo
  {author} {\bibfnamefont {M.~A.}\ \bibnamefont {Anisimov}}, \bibinfo {author}
  {\bibfnamefont {F.}~\bibnamefont {Caupin}}, \bibinfo {author} {\bibfnamefont
  {C.}~\bibnamefont {Chakravarty}}, \bibinfo {author} {\bibfnamefont
  {E.}~\bibnamefont {Lascaris}}, \bibinfo {author} {\bibfnamefont
  {T.}~\bibnamefont {Loerting}}, \bibinfo {author} {\bibfnamefont {A.~Z.}\
  \bibnamefont {Panagiotopoulos}}, \bibinfo {author} {\bibfnamefont
  {J.}~\bibnamefont {Russo}}, \bibinfo {author} {\bibfnamefont {J.~A.}\
  \bibnamefont {Sellberg}}, \bibinfo {author} {\bibfnamefont {H.~E.}\
  \bibnamefont {Stanley}}, \bibinfo {author} {\bibfnamefont {H.}~\bibnamefont
  {Tanaka}}, \bibinfo {author} {\bibfnamefont {C.}~\bibnamefont {Vega}},
  \bibinfo {author} {\bibfnamefont {L.}~\bibnamefont {Xu}}, \ and\ \bibinfo
  {author} {\bibfnamefont {L.~G.~M.}\ \bibnamefont {Pettersson}},\ }\href
  {\doibase 10.1021/acs.chemrev.5b00750} {\bibfield  {journal} {\bibinfo
  {journal} {Chem. Rev.}\ }\textbf {\bibinfo {volume} {116}},\ \bibinfo {pages}
  {7463} (\bibinfo {year} {2016})}\BibitemShut {NoStop}%
\bibitem [{\citenamefont {Cerdeiri{\~n}a}\ \emph {et~al.}(2019)\citenamefont
  {Cerdeiri{\~n}a}, \citenamefont {Troncoso}, \citenamefont
  {Gonz{\'a}lez-Salgado}, \citenamefont {Debenedetti},\ and\ \citenamefont
  {Stanley}}]{Cerdeirina2019}%
  \BibitemOpen
  \bibfield  {author} {\bibinfo {author} {\bibfnamefont {C.~A.}\ \bibnamefont
  {Cerdeiri{\~n}a}}, \bibinfo {author} {\bibfnamefont {J.}~\bibnamefont
  {Troncoso}}, \bibinfo {author} {\bibfnamefont {D.}~\bibnamefont
  {Gonz{\'a}lez-Salgado}}, \bibinfo {author} {\bibfnamefont {P.~G.}\
  \bibnamefont {Debenedetti}}, \ and\ \bibinfo {author} {\bibfnamefont {H.~E.}\
  \bibnamefont {Stanley}},\ }\href {\doibase 10.1063/1.5096890} {\bibfield
  {journal} {\bibinfo  {journal} {J. Chem. Phys.}\ }\textbf {\bibinfo {volume}
  {150}},\ \bibinfo {pages} {244509} (\bibinfo {year} {2019})}\BibitemShut
  {NoStop}%
\bibitem [{\citenamefont {Hestand}\ and\ \citenamefont
  {Skinner}(2018)}]{Hestand2018}%
  \BibitemOpen
  \bibfield  {author} {\bibinfo {author} {\bibfnamefont {N.~J.}\ \bibnamefont
  {Hestand}}\ and\ \bibinfo {author} {\bibfnamefont {J.~L.}\ \bibnamefont
  {Skinner}},\ }\href {\doibase 10.1063/1.5046687} {\bibfield  {journal}
  {\bibinfo  {journal} {J. Chem. Phys.}\ }\textbf {\bibinfo {volume} {149}},\
  \bibinfo {pages} {140901} (\bibinfo {year} {2018})}\BibitemShut {NoStop}%
\bibitem [{\citenamefont {Naserifar}\ and\ \citenamefont
  {Goddard}(2019)}]{Naserifar2019}%
  \BibitemOpen
  \bibfield  {author} {\bibinfo {author} {\bibfnamefont {S.}~\bibnamefont
  {Naserifar}}\ and\ \bibinfo {author} {\bibfnamefont {W.~A.}\ \bibnamefont
  {Goddard}},\ }\href {\doibase 10.1021/acs.jpclett.9b02443} {\bibfield
  {journal} {\bibinfo  {journal} {J. Phys. Chem. Lett.}\ }\textbf {\bibinfo
  {volume} {10}},\ \bibinfo {pages} {6267} (\bibinfo {year}
  {2019})}\BibitemShut {NoStop}%
\bibitem [{\citenamefont {Limmer}\ and\ \citenamefont
  {Chandler}(2011)}]{Limmer2011}%
  \BibitemOpen
  \bibfield  {author} {\bibinfo {author} {\bibfnamefont {D.~T.}\ \bibnamefont
  {Limmer}}\ and\ \bibinfo {author} {\bibfnamefont {D.}~\bibnamefont
  {Chandler}},\ }\href {http://link.aip.org/link/?JCP/135/134503/1} {\bibfield
  {journal} {\bibinfo  {journal} {J. Chem. Phys.}\ }\textbf {\bibinfo {volume}
  {135}},\ \bibinfo {pages} {134503} (\bibinfo {year} {2011})}\BibitemShut
  {NoStop}%
\bibitem [{\citenamefont {Limmer}\ and\ \citenamefont
  {Chandler}(2013)}]{Limmer2013}%
  \BibitemOpen
  \bibfield  {author} {\bibinfo {author} {\bibfnamefont {D.~T.}\ \bibnamefont
  {Limmer}}\ and\ \bibinfo {author} {\bibfnamefont {D.}~\bibnamefont
  {Chandler}},\ }\href {http://dx.doi.org/10.1063/1.4807479} {\bibfield
  {journal} {\bibinfo  {journal} {J. Chem. Phys.}\ }\textbf {\bibinfo {volume}
  {138}},\ \bibinfo {pages} {214504} (\bibinfo {year} {2013})}\BibitemShut
  {NoStop}%
\bibitem [{\citenamefont {Palmer}\ \emph {et~al.}(2018)\citenamefont {Palmer},
  \citenamefont {Haji-Akbari}, \citenamefont {Singh}, \citenamefont {Martelli},
  \citenamefont {Car}, \citenamefont {Panagiotopoulos},\ and\ \citenamefont
  {Debenedetti}}]{Palmer2018}%
  \BibitemOpen
  \bibfield  {author} {\bibinfo {author} {\bibfnamefont {J.~C.}\ \bibnamefont
  {Palmer}}, \bibinfo {author} {\bibfnamefont {A.}~\bibnamefont {Haji-Akbari}},
  \bibinfo {author} {\bibfnamefont {R.~S.}\ \bibnamefont {Singh}}, \bibinfo
  {author} {\bibfnamefont {F.}~\bibnamefont {Martelli}}, \bibinfo {author}
  {\bibfnamefont {R.}~\bibnamefont {Car}}, \bibinfo {author} {\bibfnamefont
  {A.~Z.}\ \bibnamefont {Panagiotopoulos}}, \ and\ \bibinfo {author}
  {\bibfnamefont {P.~G.}\ \bibnamefont {Debenedetti}},\ }\href {\doibase
  10.1063/1.5029463} {\bibfield  {journal} {\bibinfo  {journal} {J. Chem.
  Phys.}\ }\textbf {\bibinfo {volume} {148}},\ \bibinfo {pages} {137101}
  (\bibinfo {year} {2018})}\BibitemShut {NoStop}%
\bibitem [{\citenamefont {Ito}\ \emph {et~al.}(1999)\citenamefont {Ito},
  \citenamefont {Moynihan},\ and\ \citenamefont {Angell}}]{Ito1999}%
  \BibitemOpen
  \bibfield  {author} {\bibinfo {author} {\bibfnamefont {K.}~\bibnamefont
  {Ito}}, \bibinfo {author} {\bibfnamefont {C.~T.}\ \bibnamefont {Moynihan}}, \
  and\ \bibinfo {author} {\bibfnamefont {C.~A.}\ \bibnamefont {Angell}},\
  }\href {https://doi.org/10.1038/19042} {\bibfield  {journal} {\bibinfo
  {journal} {Nature}\ }\textbf {\bibinfo {volume} {398}},\ \bibinfo {pages}
  {492} (\bibinfo {year} {1999})}\BibitemShut {NoStop}%
\bibitem [{\citenamefont {Shi}\ \emph {et~al.}(2018)\citenamefont {Shi},
  \citenamefont {Russo},\ and\ \citenamefont {Tanaka}}]{Shi2018}%
  \BibitemOpen
  \bibfield  {author} {\bibinfo {author} {\bibfnamefont {R.}~\bibnamefont
  {Shi}}, \bibinfo {author} {\bibfnamefont {J.}~\bibnamefont {Russo}}, \ and\
  \bibinfo {author} {\bibfnamefont {H.}~\bibnamefont {Tanaka}},\ }\href
  {http://www.pnas.org/content/115/38/9444.abstract} {\bibfield  {journal}
  {\bibinfo  {journal} {Proc Natl Acad Sci USA}\ }\textbf {\bibinfo {volume}
  {115}},\ \bibinfo {pages} {9444} (\bibinfo {year} {2018})}\BibitemShut
  {NoStop}%
\bibitem [{\citenamefont {Dehaoui}\ \emph {et~al.}(2015)\citenamefont
  {Dehaoui}, \citenamefont {Issenmann},\ and\ \citenamefont
  {Caupin}}]{Dehaoui2015}%
  \BibitemOpen
  \bibfield  {author} {\bibinfo {author} {\bibfnamefont {A.}~\bibnamefont
  {Dehaoui}}, \bibinfo {author} {\bibfnamefont {B.}~\bibnamefont {Issenmann}},
  \ and\ \bibinfo {author} {\bibfnamefont {F.}~\bibnamefont {Caupin}},\ }\href
  {http://www.pnas.org/content/112/39/12020.abstract} {\bibfield  {journal}
  {\bibinfo  {journal} {Proc Natl Acad Sci USA}\ }\textbf {\bibinfo {volume}
  {112}},\ \bibinfo {pages} {12020} (\bibinfo {year} {2015})}\BibitemShut
  {NoStop}%
\bibitem [{\citenamefont {Singh}\ \emph {et~al.}(2017)\citenamefont {Singh},
  \citenamefont {Issenmann},\ and\ \citenamefont {Caupin}}]{Singh2017}%
  \BibitemOpen
  \bibfield  {author} {\bibinfo {author} {\bibfnamefont {L.~P.}\ \bibnamefont
  {Singh}}, \bibinfo {author} {\bibfnamefont {B.}~\bibnamefont {Issenmann}}, \
  and\ \bibinfo {author} {\bibfnamefont {F.}~\bibnamefont {Caupin}},\ }\href
  {http://www.pnas.org/content/114/17/4312.abstract} {\bibfield  {journal}
  {\bibinfo  {journal} {Proc Natl Acad Sci USA}\ }\textbf {\bibinfo {volume}
  {114}},\ \bibinfo {pages} {4312} (\bibinfo {year} {2017})}\BibitemShut
  {NoStop}%
\bibitem [{\citenamefont {Lakes}(2009)}]{Lakes2009}%
  \BibitemOpen
  \bibfield  {author} {\bibinfo {author} {\bibfnamefont {R.}~\bibnamefont
  {Lakes}},\ }\href {https://books.google.com.br/books?id=BH6f2hWWBkAC} {\emph
  {\bibinfo {title} {Viscoelastic Materials}}}\ (\bibinfo  {publisher}
  {Cambridge University Press},\ \bibinfo {year} {2009})\BibitemShut {NoStop}%
\bibitem [{\citenamefont {Krishnan}\ \emph {et~al.}(2010)\citenamefont
  {Krishnan}, \citenamefont {Deshpande},\ and\ \citenamefont
  {Kumar}}]{Krishnan2010}%
  \BibitemOpen
  \bibfield  {author} {\bibinfo {author} {\bibfnamefont {J.~M.}\ \bibnamefont
  {Krishnan}}, \bibinfo {author} {\bibfnamefont {A.~P.}\ \bibnamefont
  {Deshpande}}, \ and\ \bibinfo {author} {\bibfnamefont {P.~S.}\ \bibnamefont
  {Kumar}},\ }\href@noop {} {\emph {\bibinfo {title} {Rheology of complex
  fluids}}}\ (\bibinfo  {publisher} {Springer},\ \bibinfo {year}
  {2010})\BibitemShut {NoStop}%
\bibitem [{\citenamefont {Larson}(1999)}]{Larson1999}%
  \BibitemOpen
  \bibfield  {author} {\bibinfo {author} {\bibfnamefont {R.~G.}\ \bibnamefont
  {Larson}},\ }\href@noop {} {\emph {\bibinfo {title} {The structure and
  rheology of complex fluids}}},\ Vol.\ \bibinfo {volume} {150}\ (\bibinfo
  {publisher} {Oxford university press New York},\ \bibinfo {year}
  {1999})\BibitemShut {NoStop}%
\bibitem [{\citenamefont {Khatibi}\ \emph {et~al.}(2018)\citenamefont
  {Khatibi}, \citenamefont {Time},\ and\ \citenamefont {Shaibu}}]{Khatibi2018}%
  \BibitemOpen
  \bibfield  {author} {\bibinfo {author} {\bibfnamefont {M.}~\bibnamefont
  {Khatibi}}, \bibinfo {author} {\bibfnamefont {R.~W.}\ \bibnamefont {Time}}, \
  and\ \bibinfo {author} {\bibfnamefont {R.}~\bibnamefont {Shaibu}},\ }\href
  {http://www.sciencedirect.com/science/article/pii/S0301932217304706}
  {\bibfield  {journal} {\bibinfo  {journal} {Int. J. Multiphase Flow}\
  }\textbf {\bibinfo {volume} {99}},\ \bibinfo {pages} {284} (\bibinfo {year}
  {2018})}\BibitemShut {NoStop}%
\bibitem [{\citenamefont {Hallett}(1963)}]{Hallett1963}%
  \BibitemOpen
  \bibfield  {author} {\bibinfo {author} {\bibfnamefont {J.}~\bibnamefont
  {Hallett}},\ }\href {http://dx.doi.org/10.1088/0370-1328/82/6/326} {\bibfield
   {journal} {\bibinfo  {journal} {Proc. Phys. Soc.}\ }\textbf {\bibinfo
  {volume} {82}},\ \bibinfo {pages} {1046} (\bibinfo {year}
  {1963})}\BibitemShut {NoStop}%
\bibitem [{\citenamefont {Abascal}\ \emph {et~al.}(2005)\citenamefont
  {Abascal}, \citenamefont {Sanz}, \citenamefont {Garc{\'i}a~Fern{\'a}ndez},\
  and\ \citenamefont {Vega}}]{Abascal2005}%
  \BibitemOpen
  \bibfield  {author} {\bibinfo {author} {\bibfnamefont {J.~L.~F.}\
  \bibnamefont {Abascal}}, \bibinfo {author} {\bibfnamefont {E.}~\bibnamefont
  {Sanz}}, \bibinfo {author} {\bibfnamefont {R.}~\bibnamefont
  {Garc{\'i}a~Fern{\'a}ndez}}, \ and\ \bibinfo {author} {\bibfnamefont
  {C.}~\bibnamefont {Vega}},\ }\href {\doibase
  http://dx.doi.org/10.1063/1.1931662} {\bibfield  {journal} {\bibinfo
  {journal} {J. Chem. Phys.}\ }\textbf {\bibinfo {volume} {122}},\ \bibinfo
  {pages} {234511} (\bibinfo {year} {2005})}\BibitemShut {NoStop}%
\bibitem [{\citenamefont {Haji-Akbari}\ and\ \citenamefont
  {Debenedetti}(2017)}]{Haji-Akbari2017}%
  \BibitemOpen
  \bibfield  {author} {\bibinfo {author} {\bibfnamefont {A.}~\bibnamefont
  {Haji-Akbari}}\ and\ \bibinfo {author} {\bibfnamefont {P.~G.}\ \bibnamefont
  {Debenedetti}},\ }\href
  {http://www.pnas.org/content/early/2017/03/13/1620999114.abstract} {\bibfield
   {journal} {\bibinfo  {journal} {Proc Natl Acad Sci USA}\ ,\ \bibinfo {pages}
  {201620999}} (\bibinfo {year} {2017})}\BibitemShut {NoStop}%
\bibitem [{\citenamefont {Plimpton}(1995)}]{Plimpton1995}%
  \BibitemOpen
  \bibfield  {author} {\bibinfo {author} {\bibfnamefont {S.}~\bibnamefont
  {Plimpton}},\ }\href
  {http://www.sciencedirect.com/science/article/B6WHY-45NJN1B-3N/2/58aa2a309d2ebbbe60e0f417d398b0ef}
  {\bibfield  {journal} {\bibinfo  {journal} {J. Comput. Phys.}\ }\textbf
  {\bibinfo {volume} {117}},\ \bibinfo {pages} {1} (\bibinfo {year}
  {1995})}\BibitemShut {NoStop}%
\bibitem [{\citenamefont {Hockney}(2017)}]{Hockney2017}%
  \BibitemOpen
  \bibfield  {author} {\bibinfo {author} {\bibfnamefont {R.~W.}\ \bibnamefont
  {Hockney}},\ }\href {https://books.google.com.br/books?id=dt9-tAEACAAJ}
  {\emph {\bibinfo {title} {Computer Simulation Using Particles}}}\ (\bibinfo
  {publisher} {Taylor \& Francis Group},\ \bibinfo {year} {2017})\BibitemShut
  {NoStop}%
\bibitem [{\citenamefont {Ryckaert}\ \emph {et~al.}(1977)\citenamefont
  {Ryckaert}, \citenamefont {Ciccotti},\ and\ \citenamefont
  {Berendsen}}]{Ryckaert1977}%
  \BibitemOpen
  \bibfield  {author} {\bibinfo {author} {\bibfnamefont {J.-P.}\ \bibnamefont
  {Ryckaert}}, \bibinfo {author} {\bibfnamefont {G.}~\bibnamefont {Ciccotti}},
  \ and\ \bibinfo {author} {\bibfnamefont {H.~J.}\ \bibnamefont {Berendsen}},\
  }\href {http://www.sciencedirect.com/science/article/pii/0021999177900985}
  {\bibfield  {journal} {\bibinfo  {journal} {J. Comput. Phys.}\ }\textbf
  {\bibinfo {volume} {23}},\ \bibinfo {pages} {327} (\bibinfo {year}
  {1977})}\BibitemShut {NoStop}%
\bibitem [{\citenamefont {Shinoda}\ \emph {et~al.}(2004)\citenamefont
  {Shinoda}, \citenamefont {Shiga},\ and\ \citenamefont
  {Mikami}}]{Shinoda2004}%
  \BibitemOpen
  \bibfield  {author} {\bibinfo {author} {\bibfnamefont {W.}~\bibnamefont
  {Shinoda}}, \bibinfo {author} {\bibfnamefont {M.}~\bibnamefont {Shiga}}, \
  and\ \bibinfo {author} {\bibfnamefont {M.}~\bibnamefont {Mikami}},\ }\href
  {https://link.aps.org/doi/10.1103/PhysRevB.69.134103} {\bibfield  {journal}
  {\bibinfo  {journal} {Phys. Rev. B}\ }\textbf {\bibinfo {volume} {69}},\
  \bibinfo {pages} {134103} (\bibinfo {year} {2004})}\BibitemShut {NoStop}%
\bibitem [{\citenamefont {Schneider}\ and\ \citenamefont
  {Stoll}(1978)}]{Schneider1978}%
  \BibitemOpen
  \bibfield  {author} {\bibinfo {author} {\bibfnamefont {T.}~\bibnamefont
  {Schneider}}\ and\ \bibinfo {author} {\bibfnamefont {E.}~\bibnamefont
  {Stoll}},\ }\href {https://link.aps.org/doi/10.1103/PhysRevB.17.1302}
  {\bibfield  {journal} {\bibinfo  {journal} {Phys. Rev. B}\ }\textbf {\bibinfo
  {volume} {17}},\ \bibinfo {pages} {1302} (\bibinfo {year}
  {1978})}\BibitemShut {NoStop}%
\bibitem [{\citenamefont {Hagita}\ \emph {et~al.}(2017)\citenamefont {Hagita},
  \citenamefont {Murashima}, \citenamefont {Takano},\ and\ \citenamefont
  {Kawakatsu}}]{Hagita2017}%
  \BibitemOpen
  \bibfield  {author} {\bibinfo {author} {\bibfnamefont {K.}~\bibnamefont
  {Hagita}}, \bibinfo {author} {\bibfnamefont {T.}~\bibnamefont {Murashima}},
  \bibinfo {author} {\bibfnamefont {H.}~\bibnamefont {Takano}}, \ and\ \bibinfo
  {author} {\bibfnamefont {T.}~\bibnamefont {Kawakatsu}},\ }\href {\doibase
  10.7566/JPSJ.86.124803} {\bibfield  {journal} {\bibinfo  {journal} {J. Phys.
  Soc. Jpn.}\ }\textbf {\bibinfo {volume} {86}},\ \bibinfo {pages} {124803}
  (\bibinfo {year} {2017})}\BibitemShut {NoStop}%
\bibitem [{\citenamefont {Osaki}\ \emph {et~al.}(2000)\citenamefont {Osaki},
  \citenamefont {Inoue},\ and\ \citenamefont {Isomura}}]{Osaki2000}%
  \BibitemOpen
  \bibfield  {author} {\bibinfo {author} {\bibfnamefont {K.}~\bibnamefont
  {Osaki}}, \bibinfo {author} {\bibfnamefont {T.}~\bibnamefont {Inoue}}, \ and\
  \bibinfo {author} {\bibfnamefont {T.}~\bibnamefont {Isomura}},\ }\href
  {\doibase 10.1002/1099-0488(20000715)38:14<1917::AID-POLB100>3.0.CO;2-6}
  {\bibfield  {journal} {\bibinfo  {journal} {J. Polym. Sci. B Polym. Phys.}\
  }\textbf {\bibinfo {volume} {38}},\ \bibinfo {pages} {1917} (\bibinfo {year}
  {2000})}\BibitemShut {NoStop}%
\bibitem [{\citenamefont {Islam}\ and\ \citenamefont
  {Archer}(2001)}]{Islam2001}%
  \BibitemOpen
  \bibfield  {author} {\bibinfo {author} {\bibfnamefont {M.~T.}\ \bibnamefont
  {Islam}}\ and\ \bibinfo {author} {\bibfnamefont {L.~A.}\ \bibnamefont
  {Archer}},\ }\href {\doibase 10.1002/polb.1201} {\bibfield  {journal}
  {\bibinfo  {journal} {J. Polym. Sci. B Polym. Phys.}\ }\textbf {\bibinfo
  {volume} {39}},\ \bibinfo {pages} {2275} (\bibinfo {year}
  {2001})}\BibitemShut {NoStop}%
\bibitem [{\citenamefont {Letwimolnun}\ \emph {et~al.}(2007)\citenamefont
  {Letwimolnun}, \citenamefont {Vergnes}, \citenamefont {Ausias},\ and\
  \citenamefont {Carreau}}]{Letwimolnun2007}%
  \BibitemOpen
  \bibfield  {author} {\bibinfo {author} {\bibfnamefont {W.}~\bibnamefont
  {Letwimolnun}}, \bibinfo {author} {\bibfnamefont {B.}~\bibnamefont
  {Vergnes}}, \bibinfo {author} {\bibfnamefont {G.}~\bibnamefont {Ausias}}, \
  and\ \bibinfo {author} {\bibfnamefont {P.~J.}\ \bibnamefont {Carreau}},\
  }\href {http://www.sciencedirect.com/science/article/pii/S0377025706002953}
  {\bibfield  {journal} {\bibinfo  {journal} {J. Non-Newtonian Fluid Mech.}\
  }\textbf {\bibinfo {volume} {141}},\ \bibinfo {pages} {167} (\bibinfo {year}
  {2007})}\BibitemShut {NoStop}%
\bibitem [{\citenamefont {Varnik}\ \emph {et~al.}(2004)\citenamefont {Varnik},
  \citenamefont {Bocquet},\ and\ \citenamefont {Barrat}}]{Varnik2004}%
  \BibitemOpen
  \bibfield  {author} {\bibinfo {author} {\bibfnamefont {F.}~\bibnamefont
  {Varnik}}, \bibinfo {author} {\bibfnamefont {L.}~\bibnamefont {Bocquet}}, \
  and\ \bibinfo {author} {\bibfnamefont {J.-L.}\ \bibnamefont {Barrat}},\
  }\href {\doibase 10.1063/1.1636451} {\bibfield  {journal} {\bibinfo
  {journal} {J. Chem. Phys.}\ }\textbf {\bibinfo {volume} {120}},\ \bibinfo
  {pages} {2788} (\bibinfo {year} {2004})}\BibitemShut {NoStop}%
\bibitem [{\citenamefont {Zausch}\ \emph {et~al.}(2008)\citenamefont {Zausch},
  \citenamefont {Horbach}, \citenamefont {Laurati}, \citenamefont {Egelhaaf},
  \citenamefont {Brader}, \citenamefont {Voigtmann},\ and\ \citenamefont
  {Fuchs}}]{Zausch2008}%
  \BibitemOpen
  \bibfield  {author} {\bibinfo {author} {\bibfnamefont {J.}~\bibnamefont
  {Zausch}}, \bibinfo {author} {\bibfnamefont {J.}~\bibnamefont {Horbach}},
  \bibinfo {author} {\bibfnamefont {M.}~\bibnamefont {Laurati}}, \bibinfo
  {author} {\bibfnamefont {S.~U.}\ \bibnamefont {Egelhaaf}}, \bibinfo {author}
  {\bibfnamefont {J.~M.}\ \bibnamefont {Brader}}, \bibinfo {author}
  {\bibfnamefont {T.}~\bibnamefont {Voigtmann}}, \ and\ \bibinfo {author}
  {\bibfnamefont {M.}~\bibnamefont {Fuchs}},\ }\href
  {http://dx.doi.org/10.1088/0953-8984/20/40/404210} {\bibfield  {journal}
  {\bibinfo  {journal} {J. Phys.: Condens. Matter}\ }\textbf {\bibinfo {volume}
  {20}},\ \bibinfo {pages} {404210} (\bibinfo {year} {2008})}\BibitemShut
  {NoStop}%
\bibitem [{\citenamefont {Zausch}\ and\ \citenamefont
  {Horbach}(2009)}]{Zausch2009}%
  \BibitemOpen
  \bibfield  {author} {\bibinfo {author} {\bibfnamefont {J.}~\bibnamefont
  {Zausch}}\ and\ \bibinfo {author} {\bibfnamefont {J.}~\bibnamefont
  {Horbach}},\ }\href {http://dx.doi.org/10.1209/0295-5075/88/60001} {\bibfield
   {journal} {\bibinfo  {journal} {EPL}\ }\textbf {\bibinfo {volume} {88}},\
  \bibinfo {pages} {60001} (\bibinfo {year} {2009})}\BibitemShut {NoStop}%
\bibitem [{\citenamefont {Fuereder}\ and\ \citenamefont
  {Ilg}(2017)}]{Fuereder2017}%
  \BibitemOpen
  \bibfield  {author} {\bibinfo {author} {\bibfnamefont {I.}~\bibnamefont
  {Fuereder}}\ and\ \bibinfo {author} {\bibfnamefont {P.}~\bibnamefont {Ilg}},\
  }\href {http://dx.doi.org/10.1039/C7SM00178A} {\bibfield  {journal} {\bibinfo
   {journal} {Soft Matter}\ }\textbf {\bibinfo {volume} {13}},\ \bibinfo
  {pages} {2192} (\bibinfo {year} {2017})}\BibitemShut {NoStop}%
\bibitem [{\citenamefont {Carreau}(1972)}]{Carreau1972}%
  \BibitemOpen
  \bibfield  {author} {\bibinfo {author} {\bibfnamefont {P.~J.}\ \bibnamefont
  {Carreau}},\ }\href {\doibase 10.1122/1.549276} {\bibfield  {journal}
  {\bibinfo  {journal} {Trans. Soc. Rheol.}\ }\textbf {\bibinfo {volume}
  {16}},\ \bibinfo {pages} {99} (\bibinfo {year} {1972})}\BibitemShut {NoStop}%
\bibitem [{\citenamefont {Spikes}\ and\ \citenamefont
  {Jie}(2014)}]{Spikes2014}%
  \BibitemOpen
  \bibfield  {author} {\bibinfo {author} {\bibfnamefont {H.}~\bibnamefont
  {Spikes}}\ and\ \bibinfo {author} {\bibfnamefont {Z.}~\bibnamefont {Jie}},\
  }\href {https://doi.org/10.1007/s11249-014-0396-y} {\bibfield  {journal}
  {\bibinfo  {journal} {Tribol. Lett.}\ }\textbf {\bibinfo {volume} {56}},\
  \bibinfo {pages} {1} (\bibinfo {year} {2014})}\BibitemShut {NoStop}%
\bibitem [{\citenamefont {Valencia-Jaime}\ \emph {et~al.}(2019)\citenamefont
  {Valencia-Jaime}, \citenamefont {Desgranges},\ and\ \citenamefont
  {Delhommelle}}]{Valencia2019}%
  \BibitemOpen
  \bibfield  {author} {\bibinfo {author} {\bibfnamefont {I.}~\bibnamefont
  {Valencia-Jaime}}, \bibinfo {author} {\bibfnamefont {C.}~\bibnamefont
  {Desgranges}}, \ and\ \bibinfo {author} {\bibfnamefont {J.}~\bibnamefont
  {Delhommelle}},\ }\href@noop {} {\bibfield  {journal} {\bibinfo  {journal}
  {Chem. Phys. Lett.}\ }\textbf {\bibinfo {volume} {719}},\ \bibinfo {pages}
  {103} (\bibinfo {year} {2019})}\BibitemShut {NoStop}%
\bibitem [{\citenamefont {Jadhao}\ and\ \citenamefont
  {Robbins}(2017)}]{Jadhao2017}%
  \BibitemOpen
  \bibfield  {author} {\bibinfo {author} {\bibfnamefont {V.}~\bibnamefont
  {Jadhao}}\ and\ \bibinfo {author} {\bibfnamefont {M.~O.}\ \bibnamefont
  {Robbins}},\ }\href@noop {} {\bibfield  {journal} {\bibinfo  {journal} {Proc.
  Natl. Acad. Sci. U.S.A.}\ }\textbf {\bibinfo {volume} {114}},\ \bibinfo
  {pages} {7952} (\bibinfo {year} {2017})}\BibitemShut {NoStop}%
\bibitem [{\citenamefont {Hansen}\ and\ \citenamefont
  {McDonald}(2006)}]{Hansen2006}%
  \BibitemOpen
  \bibfield  {author} {\bibinfo {author} {\bibfnamefont {J.}~\bibnamefont
  {Hansen}}\ and\ \bibinfo {author} {\bibfnamefont {I.}~\bibnamefont
  {McDonald}},\ }\href {http://books.google.com/books?id=Uhm87WZBnxEC} {\emph
  {\bibinfo {title} {Theory of simple liquids}}},\ \bibinfo {edition} {3rd}\
  ed.\ (\bibinfo  {publisher} {Elsevier Academic Press},\ \bibinfo {year}
  {2006})\BibitemShut {NoStop}%
\bibitem [{\citenamefont {Allen}\ \emph {et~al.}(2017)\citenamefont {Allen},
  \citenamefont {Tildesley},\ and\ \citenamefont {Tildesley}}]{Allen2017}%
  \BibitemOpen
  \bibfield  {author} {\bibinfo {author} {\bibfnamefont {M.~P.}\ \bibnamefont
  {Allen}}, \bibinfo {author} {\bibfnamefont {D.~J.}\ \bibnamefont
  {Tildesley}}, \ and\ \bibinfo {author} {\bibfnamefont {D.~J.}\ \bibnamefont
  {Tildesley}},\ }\href {https://books.google.com.br/books?id=PA9rnQAACAAJ}
  {\emph {\bibinfo {title} {Computer Simulation of Liquids}}}\ (\bibinfo
  {publisher} {Oxford University Press},\ \bibinfo {year} {2017})\BibitemShut
  {NoStop}%
\bibitem [{\citenamefont {McQuarrie}(2000)}]{McQuarrie2000}%
  \BibitemOpen
  \bibfield  {author} {\bibinfo {author} {\bibfnamefont {D.}~\bibnamefont
  {McQuarrie}},\ }\href {http://books.google.com.br/books?id=itcpPnDnJM0C}
  {\emph {\bibinfo {title} {Statistical Mechanics}}}\ (\bibinfo  {publisher}
  {University Science Books},\ \bibinfo {year} {2000})\BibitemShut {NoStop}%
\bibitem [{\citenamefont {Alf\'{e}}\ and\ \citenamefont
  {Gillan}(1998)}]{Alfe1998}%
  \BibitemOpen
  \bibfield  {author} {\bibinfo {author} {\bibfnamefont {D.}~\bibnamefont
  {Alf\'{e}}}\ and\ \bibinfo {author} {\bibfnamefont {M.~J.}\ \bibnamefont
  {Gillan}},\ }\href {http://link.aps.org/doi/10.1103/PhysRevLett.81.5161}
  {\bibfield  {journal} {\bibinfo  {journal} {Phys. Rev. Lett.}\ }\textbf
  {\bibinfo {volume} {81}},\ \bibinfo {pages} {5161} (\bibinfo {year}
  {1998})}\BibitemShut {NoStop}%
\bibitem [{\citenamefont {Debenedetti}\ and\ \citenamefont
  {Stillinger}(2001)}]{Debenedetti2001}%
  \BibitemOpen
  \bibfield  {author} {\bibinfo {author} {\bibfnamefont {P.~G.}\ \bibnamefont
  {Debenedetti}}\ and\ \bibinfo {author} {\bibfnamefont {F.~H.}\ \bibnamefont
  {Stillinger}},\ }\href {https://doi.org/10.1038/35065704} {\bibfield
  {journal} {\bibinfo  {journal} {Nature}\ }\textbf {\bibinfo {volume} {410}},\
  \bibinfo {pages} {259} (\bibinfo {year} {2001})}\BibitemShut {NoStop}%
\bibitem [{\citenamefont {Loose}\ and\ \citenamefont {Hess}(1989)}]{Loose1989}%
  \BibitemOpen
  \bibfield  {author} {\bibinfo {author} {\bibfnamefont {W.}~\bibnamefont
  {Loose}}\ and\ \bibinfo {author} {\bibfnamefont {S.}~\bibnamefont {Hess}},\
  }\href {https://doi.org/10.1007/BF01356970} {\bibfield  {journal} {\bibinfo
  {journal} {Rheol. Acta}\ }\textbf {\bibinfo {volume} {28}},\ \bibinfo {pages}
  {91} (\bibinfo {year} {1989})}\BibitemShut {NoStop}%
\bibitem [{\citenamefont {Kr{\"o}ger}\ and\ \citenamefont
  {Hess}(1993)}]{Kroger1993}%
  \BibitemOpen
  \bibfield  {author} {\bibinfo {author} {\bibfnamefont {M.}~\bibnamefont
  {Kr{\"o}ger}}\ and\ \bibinfo {author} {\bibfnamefont {S.}~\bibnamefont
  {Hess}},\ }\href
  {http://www.sciencedirect.com/science/article/pii/037843719390162W}
  {\bibfield  {journal} {\bibinfo  {journal} {Physica A: Statistical Mechanics
  and its Applications}\ }\textbf {\bibinfo {volume} {195}},\ \bibinfo {pages}
  {336} (\bibinfo {year} {1993})}\BibitemShut {NoStop}%
\bibitem [{\citenamefont {Petravic}\ and\ \citenamefont
  {Delhommelle}(2005)}]{Petravic2005}%
  \BibitemOpen
  \bibfield  {author} {\bibinfo {author} {\bibfnamefont {J.}~\bibnamefont
  {Petravic}}\ and\ \bibinfo {author} {\bibfnamefont {J.}~\bibnamefont
  {Delhommelle}},\ }\href {\doibase 10.1063/1.1940050} {\bibfield  {journal}
  {\bibinfo  {journal} {J. Chem. Phys.}\ }\textbf {\bibinfo {volume} {122}},\
  \bibinfo {pages} {234509} (\bibinfo {year} {2005})}\BibitemShut {NoStop}%
\bibitem [{\citenamefont {Lubchenko}(2009)}]{Lubchenko2009}%
  \BibitemOpen
  \bibfield  {author} {\bibinfo {author} {\bibfnamefont {V.}~\bibnamefont
  {Lubchenko}},\ }\href {http://www.pnas.org/content/106/28/11506.abstract}
  {\bibfield  {journal} {\bibinfo  {journal} {Proc Natl Acad Sci USA}\ }\textbf
  {\bibinfo {volume} {106}},\ \bibinfo {pages} {11506} (\bibinfo {year}
  {2009})}\BibitemShut {NoStop}%
\end{thebibliography}
%

\end{document}